\documentclass{cvmp_short}

\begin{document}

\title{Generating real-time detailed ground visualisations from sparse aerial point clouds}

\addauthor{Aidan Murray, Scarlet Mitchell, Alexander Bradley}{www.cobrasimulation.com}{1}
\addauthor{Eddie Waite, Caleb Ross, Joanna Jamrozy, Kenny Mitchell}{www.napier.ac.uk}{2}

\addinstitution{
Cobra Simulation Ltd,\\
 Bathgate}
\addinstitution{
Edinburgh Napier University and Cobra Simulation Ltd,\\ Edinburgh}
\maketitle


\noindent

\newcommand{\tr}{^{\top}}
\newcommand{\matx}[1]{\mbox{\tt #1}}
\newcommand{\vect}[1]{{\bf #1}}
\section{Procedural LIDAR point cloud detail amplification}
Building realistic wide scale outdoor 3D content with sufficient visual quality to observe at walking eye level or from driven vehicles is often carried out by large teams of artists skilled in modelling, texturing, material shading and lighting, which typically leads to both prohibitive costs and reduced accuracy honoring the variety of real world ground truth landscapes. In our proposed method, we define a process to automatically amplify real-world scanned data and render real-time in animated 3D to explore at close range with high quality for training, simulation, video game and visualisation applications. 

\begin{figure}
    \centering
    \includegraphics[width=.45\textwidth]{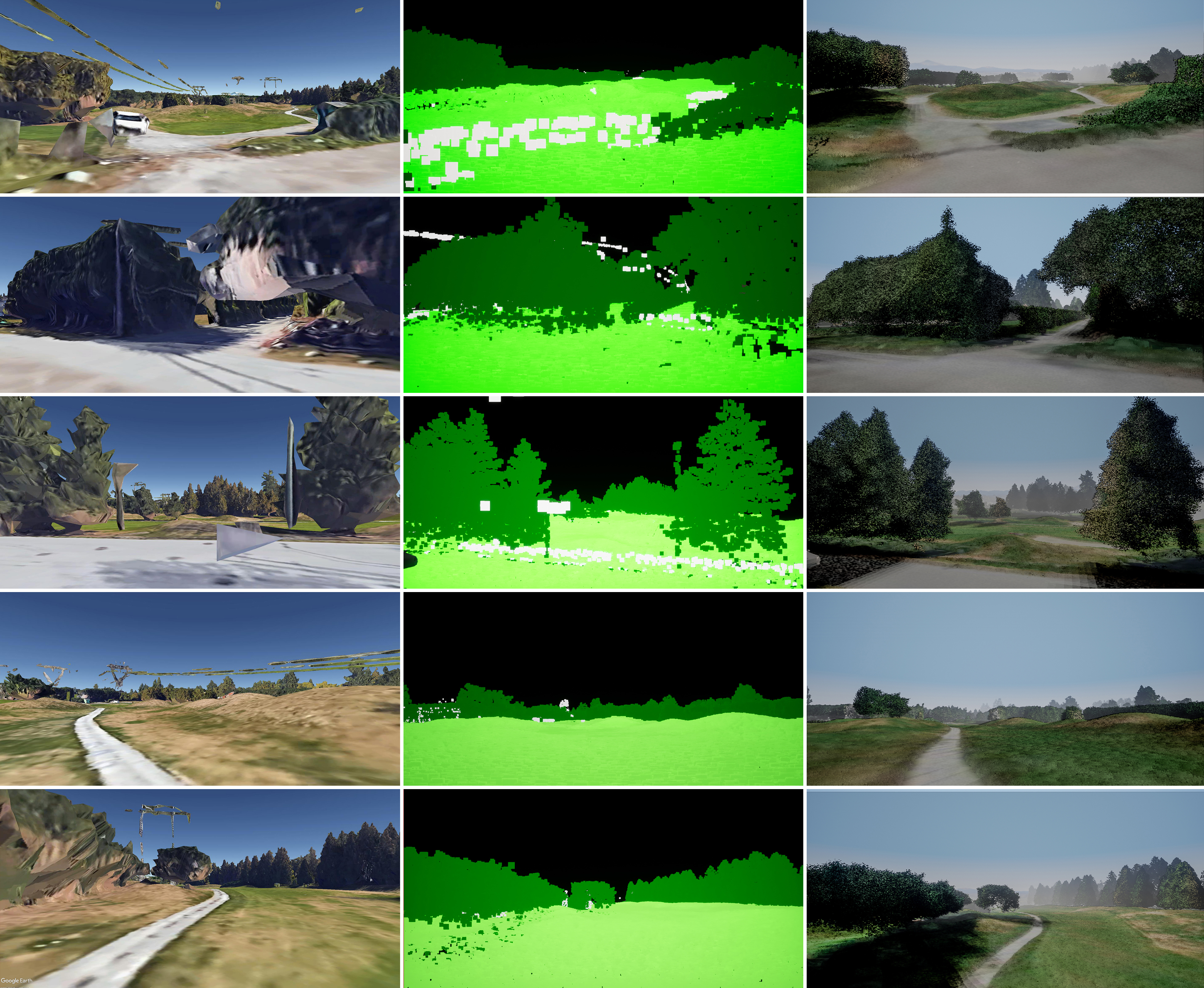}
    \caption{Comparison of similar walking eye level viewpoints, between Google Earth (left), the Epic Games' Unreal Engine 5.0 LiDAR plugin illustrating source point data (middle) and our method (right).}
    \label{fig:comparisons - Google Earth}
\end{figure}

We perform procedural 3D world generation processing upon remote sensed aerial LiDAR scan data \cite{varney2020dales}, which is often noisy, sparse and incomplete. However, given procedural, categorised shape function interpolation with signed distance fields (SDF) ray marching, the sparse data holds sufficient key information to reconstruct large areas of terrain rapidly formed within spanned volume extents, that encode nearest adjacent points, classification and material properties. That is, we infer rich local surface material, and shape details from simple point cloud data clusters by mapping matched ‘templates’ to high-quality pre-scanned local surfaces. Each template holds procedural variation so no two surfaces are exactly alike, and blend seamlessly together with noise and prior based distance field interpolation methods to combine and form a complete digital twin of the real world. We compare with Google Earth in figure \ref{fig:comparisons - Google Earth}.

A nearest relative of our work is GANCraft which takes the low fidelity blocks world representation of the video game Minecraft and generates towards realistic results with a neural rendering framework \cite{Hao2021GANcraftU3}. However, whilst our results are free of cubist style artifacts, the blocky outline appearance is not addressed by GANCraft. 

The scheme takes points of spatial scene data and generates arbitrary 3D views with high density inferred shape corresponding to the original reference scanned location through point based distance field interpolated rendering, including surface color (albedo) from registered aerial maps.

Our lean new local point based extended volume element rendering descriptor, includes cached point adjacency of neighbouring points, along wide material and shading data intended for efficiently combining on GPU for real-time rendered reconstruction of scenery.

The signed distance field (SDF) based detailed shape definition, whose volume comprises the set of adjacency vectors to neighbouring point cloud points forming, e.g., capsules emanating from a central 3D location, and may be any shape derivable from 'star'-like structured volume elements.

Detailed appearance reconstruction is achieved through authored analytic SDF shaping and noise functions according to common feature classifications, e.g. trees, grass, roads, etc. recovering spatially varying material appearance properties including diffuse, specular, roughness, etc. 

Hierarchical spatial partitioning (3D grid buckets) is performed for compute shader efficient multi-threaded data transformation of the raw sparse point cloud data into efficient fragment shader rendering packets with up to eight neighbouring adjacent point cloud data. 

We provide an efficient ray marched screen aligned quad rendering of the inferred detail SDFs with temporal re-projection depth based occlusion culling optimization, bounding sphere limit culling, and per chunk culling. Critically, to counter overdraw of naive quad rendering we perform occlusion culling with re-projection of approximate scene depth from the previous frame to sparingly cull at the level of quads.




\begin{figure}
    \centering
    \includegraphics[width=.45\textwidth]{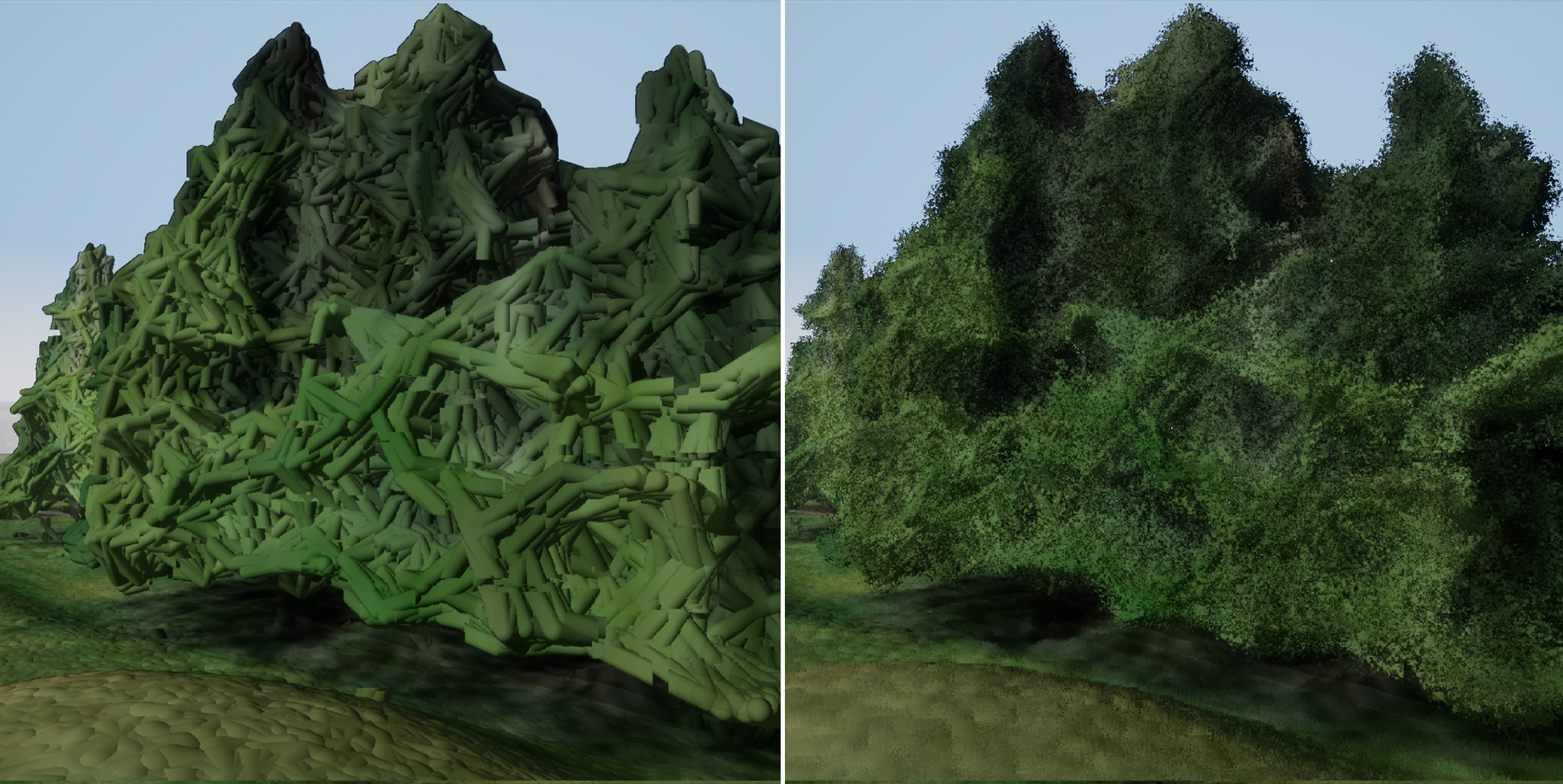}
    \caption{Tree reconstruction inferred from sparse point cloud data showing before and after addition of noise-based SDF natural shape detail.}
    \label{fig:Treenoise}
\end{figure}

\begin{figure}
  \centering
  \includegraphics[width=0.09\textwidth]{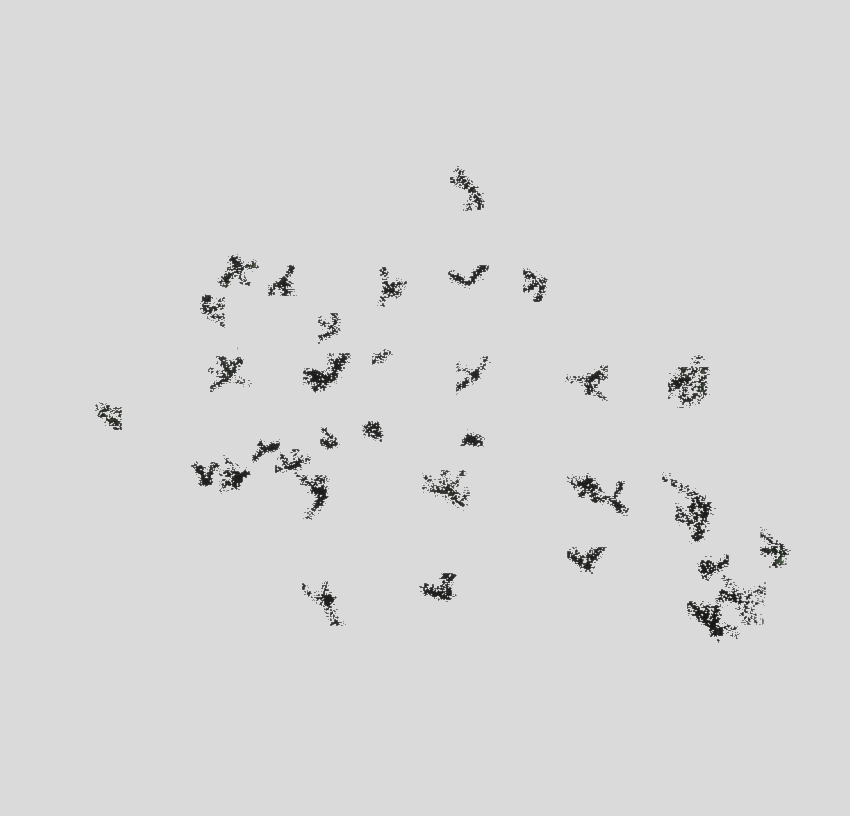}
  \includegraphics[width=0.09\textwidth]{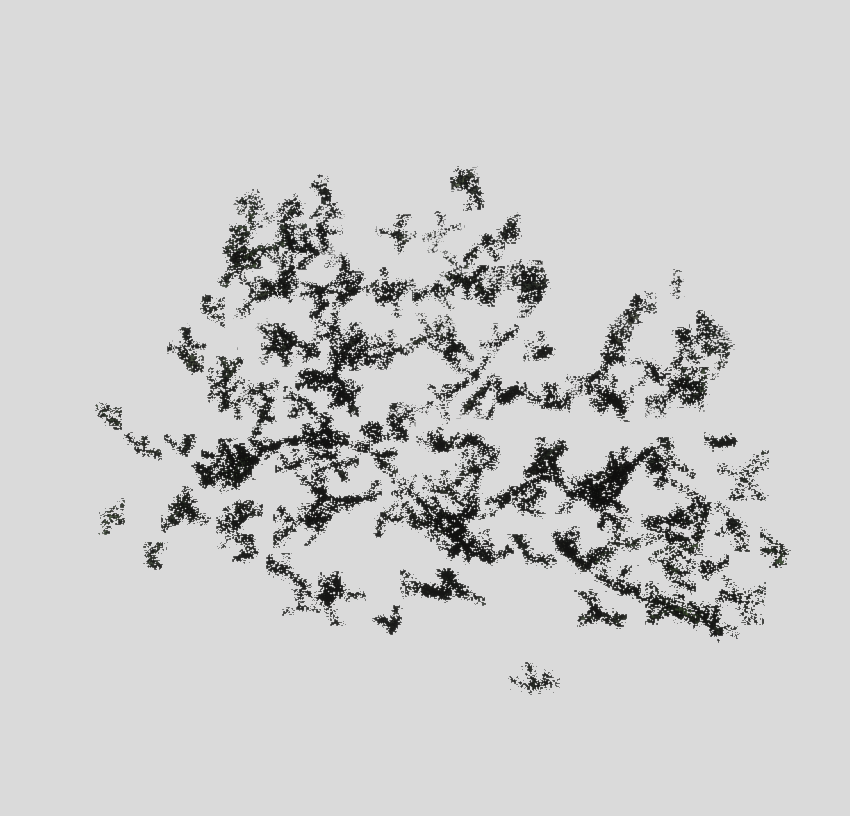}
  \includegraphics[width=0.09\textwidth]{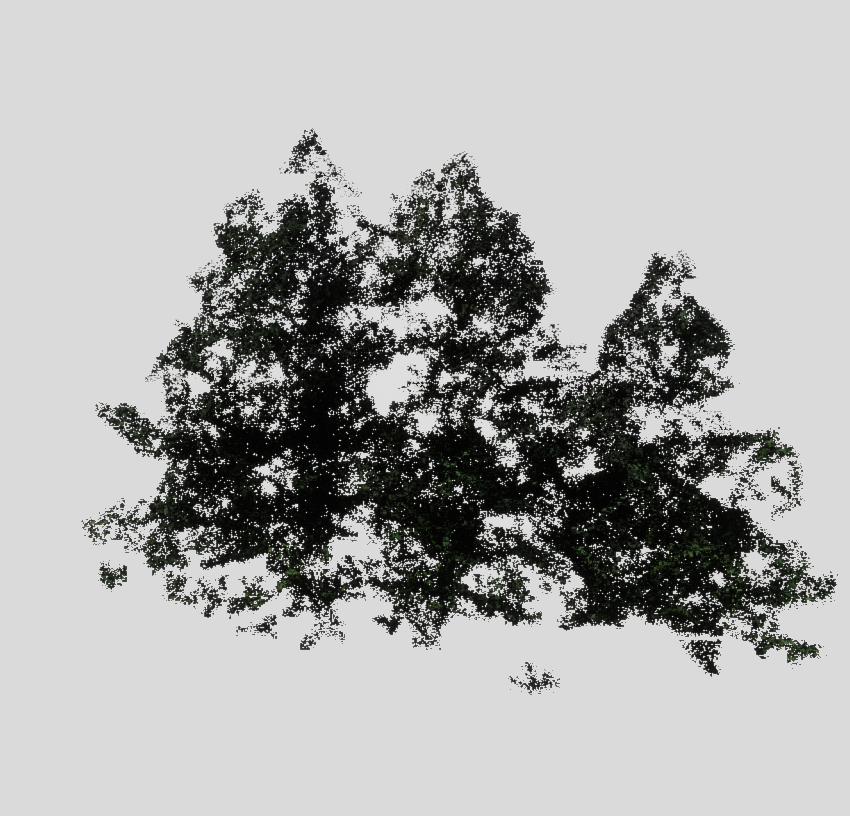}
  \includegraphics[width=0.09\textwidth]{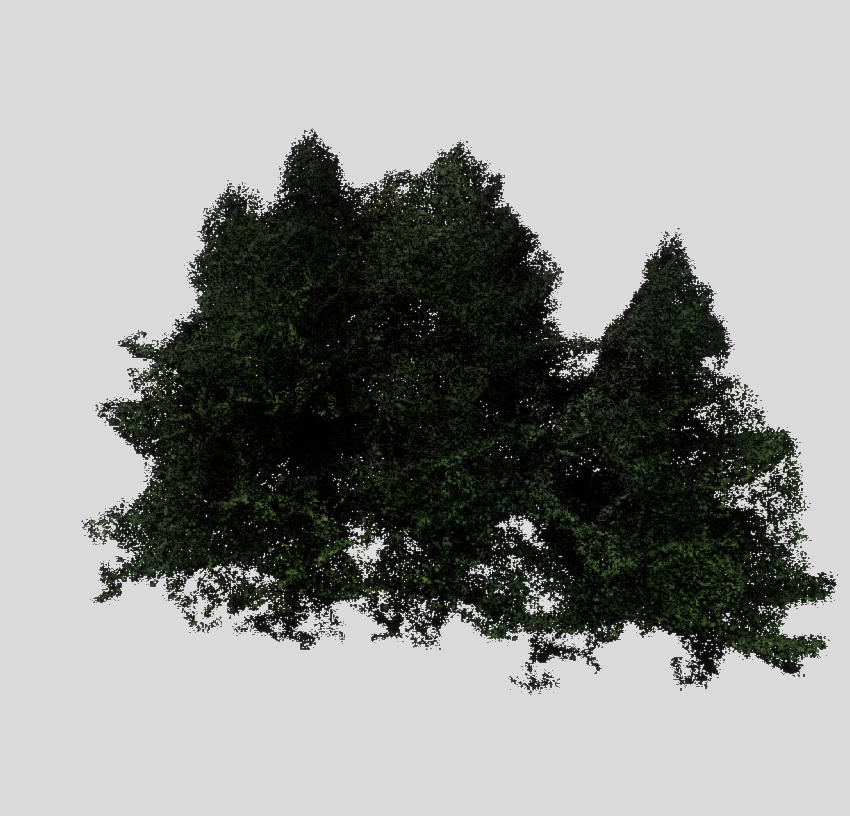}
  \includegraphics[width=0.09\textwidth]{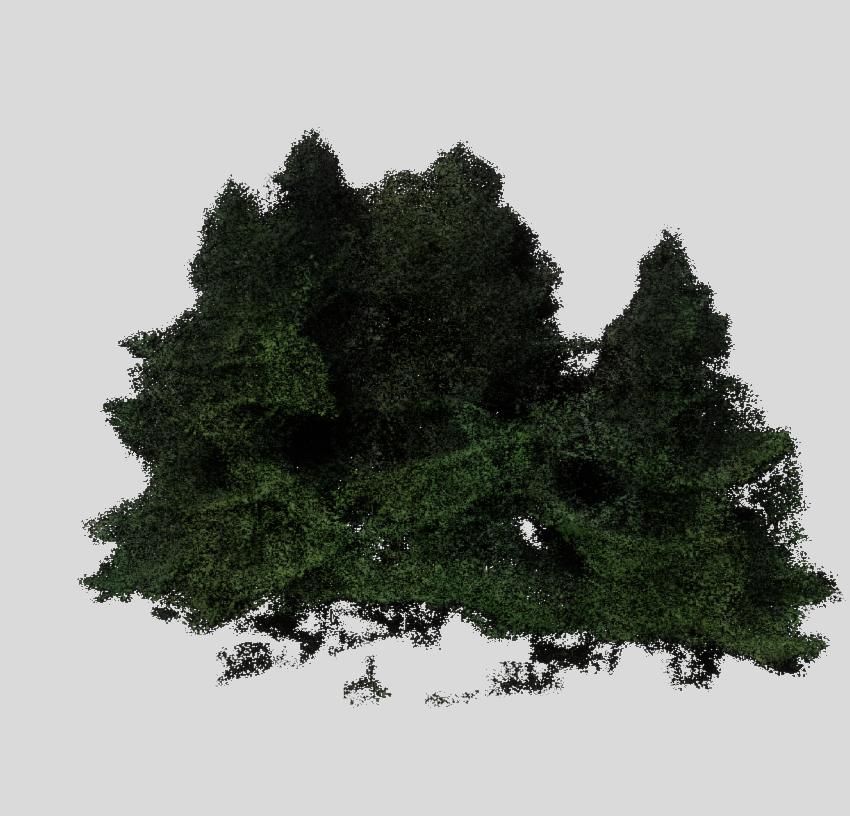}
\caption{The combination of extended point surface star-like capsules forms resulting tree structures inferred from sparse 1m resolution data.}
\label{fig:Alternative shapes}
\end{figure}

Ground level natural surfaces are achieved through offsetting approximately shaped surfaces based on increasing frequencies of directed shape function according to level of detail, e.g., a vertically tapered noise perturbed grass surface. For trees and bushes, the extended point derived surface is generated with shape steered by the connectivity information of adjacent point cloud data of same foliage category (fig. \ref{fig:Treenoise}). This adjacency data in our example forms oriented SDF capsules that, when composited, resolve to a detailed appearance of natural landscape features, e.g leaf-like features by offsetting the shape's surface in all directions (fig. \ref{fig:Alternative shapes}). 

Our new scheme provides a both texture-less and mesh-less framework, which affords memory efficient, flexible level of detail and visibility optimizations for rendering scenery in real-time. Generating rapid walk-through environments from efficiently gathered sparse LIDAR data forms a new avenue for digital twins for simulation and training, and entertainment, and provides a seed for future analytic and neural extended point-based rendering approaches.
\small{This work was supported by the European Union’s
Horizon 2020 research and innovation programme
under Grant 101017779.}
\bibliography{references}

\end{document}